# Power Flow Analysis of a 5-Bus Power System Based on Newton-Raphson Method

Sampson E. Nwachukwu, *Graduate Student Member, IEEE*

***Abstract*—Load flow analysis is a fundamental technique used by electrical engineers to simulate and evaluate power system behavior under steady-state conditions. It enables efficient operation and control by determining how active and reactive power flows throughout the system. Selecting an appropriate solution method is critical to ensuring the reliable and economical operation of power generation, transmission, and distribution networks. While the conventional loop method may be used in small-scale systems, it is limited by its reliance on impedance-based load data and its inability to scale to complex networks. In contrast, iterative techniques such as the Gauss-Seidel (GS) and Newton-Raphson (NR) methods are better suited for analyzing large systems. Of these, the NR method offers significant advantages due to its quadratic convergence and improved numerical stability. This study presents a power flow analysis of a 5-bus system using the Newton-Raphson approach. The system was modeled and simulated in PowerWorld Simulator (PWS), and a custom MATLAB implementation was developed to verify the results under a base case scenario. The comparative analysis demonstrates that the NR method provides accurate and robust solutions for power flow problems, making it well-suited for evaluating system performance under various operating conditions.**

## I. INTRODUCTION

A POWER system uses a number of network branches to transfer electricity from the generating center to the load. The term "load flow" or "power flow" refers to the flow of both active and reactive power. Load flow analysis is therefore a vital method employed by engineers to design and determine the power system's steady-state operation[1]. Additionally, a lot of other analyses, such as economic scheduling and contingency studies, require power flow analysis [2]. Under steady-state conditions, power flow studies provide a mathematical framework for estimating reactive and active power flows across different loads, phase angles, bus voltages, transformer settings, etc [1], [3]. Therefore, iterative numerical techniques are needed to solve the power flow equations, which are the resulting power-related equations that become nonlinear[1].

For power-flow studies, conventional loop or nodal analysis is unsuitable as input data for loads is not provided in terms of impedance. Instead, they are often provided in terms of power. Also, instead of being sources of voltage or current, generators are regarded as power sources [3]. Furthermore, estimating the operational parameters of a small number of individual circuits can be done by the conventional method, but without computer programs, it would be impossible to perform precise estimates of load flows or short circuit analyses in exponentially increased power systems [1]. Thus, in order to define the power flow problem, a set of computer-solvable nonlinear algebraic equations is employed [1], [3].

The most popular methods for solving power flow problems are the Newton Raphson (NR), Gauss-Seidel (GS), and fast decoupled load flow solution methods [1]–[4]. These numerical methods vary because of how these nonlinear equations are solved [4]. Generally, a faster solution time and a high level of accuracy are needed to choose the optimal numerical approach for power flow analysis [1]. In addition to being less likely to deviate from ill-conditioned problems, NR's approach is mathematically superior to other approaches due to its quadratic convergence [2]. The NR approach has been determined to be more effective and feasible for large power systems. The scale of the system has little bearing on how many iterations are needed to find a solution, but each iteration requires additional operational assessments. The power flow equation is defined in polar form since the voltage-controlled buses in the power flow problem have voltage magnitude and real power defined [2].

As a result, this study presents an NR flow analysis of a 5-bus power system. The system is modeled and simulated for 50 iterations using the PowerWorld Simulator. The goal is to analyze the system in different operating conditions, including normal operation, transformer tap changing condition, and different faults (e.g., line-line, single line to ground, double-line to ground, and three phase to ground faults) that occur at the system bus. The fault analysis helps to determine the capacity of the circuit breakers required to protect the system in real-life scenarios. Also, a MATLAB code was developed to compare the results obtained using the PowerWorld Simulator (PWS). The results obtained show that the numerical approach, such as NR for power flow analysis is more suitable and accurate for determining the behavior of the system in different operating conditions.

The remainder of this project is organized as follows: Section II presents the mathematical formulation of the power flow problem. Section III presents the design methodology and system modeling. Section IV presents the simulation results and discussion. Section V presents the conclusion of the study.

## II. FORMULATION OF THE POWER FLOW PROBLEM

A single-line diagram of the power system serves as the foundation for a power-flow problem and provides the input data for computer solutions. Here, data from transformers, transmission lines, and buses are all considered input data [2], [3].

### A. Categorization of Bus

A bus is a node that has one or more connections for generators, transmission lines, and loads [1]. Four parameters belong to each bus $k$, as illustrated in Fig.1: the bus's phase angle ($\delta_k$), voltage magnitude ($V_k$), reactive power ($Q_k$), and net real power ($P_k$). Two of these parameters are supplied as input data at each bus, while the other two are unknowns that the PWS must determine using the parameters that are known [1]–[3].

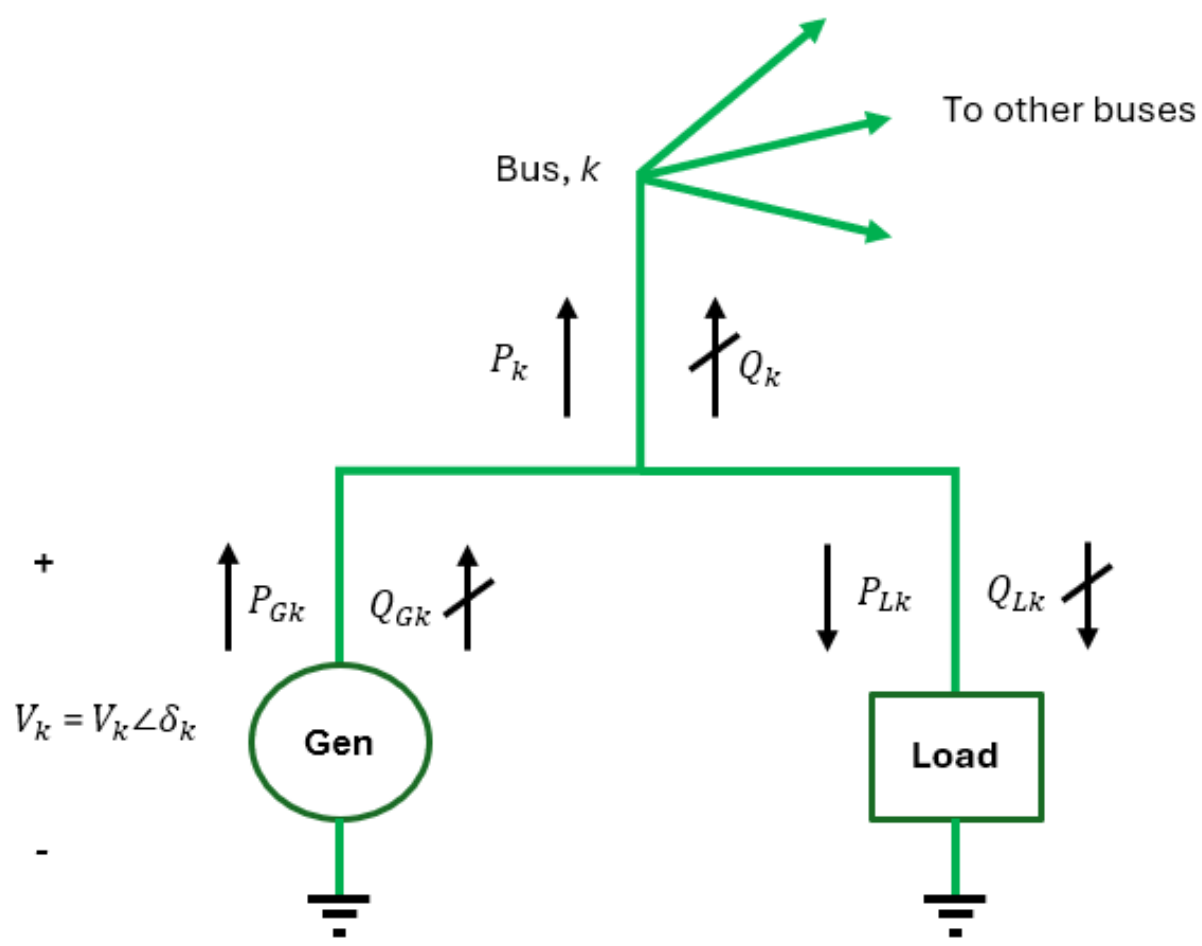

Fig. 1. Diagram showing the bus variables [3]

The power supplied to bus $k$ in Fig.1 is divided into generator and load quantities for simplicity, as expressed in (1) [1]–[3]. The following subsections present the categories of buses.

$$\begin{aligned} P_k &= P_{G,k} - P_{L,k} \\ Q_k &= Q_{G,k} - Q_{L,k} \end{aligned} \tag{1}$$

#### 1) Swing bus (or Slack bus)

The power balancing criteria are met by using this as a reference/slack bus. In a power system, bus 1 is typically used to identify the slack bus. On this bus, $P_k$ and $Q_k$ are unknown variables, but $V_k$ and $\delta_k$ are known variables. Typically, the input data, $V_1\angle\delta_1$ is represented as $1.0\angle 0^o$ per unit (pu) [1]–[3].

#### 2) Generator Bus (PV)

This bus is used for voltage regulation. Connected to a generator unit, the bus's output power may be adjusted by varying the prime mover, and the generator's excitation can be adjusted to change the voltage. This bus's unknown variables are $Q_k$ and $\delta_k$, whereas its known variables are $P_k$ and $V_k$ [1]–[3].

#### 3) Load (PQ) Bus

Measurements or historical data records can be used to determine this bus, which is not a generator. The power consumption in a power system is characterized as negative, whereas the real and reactive power supply to a power system is defined as positive. This bus meets the needs of the consumer. Here, $P_k$ and $Q_k$ are known variables for this bus, but $V_k$ and $\delta_k$ are unknown variables [1]–[3]. The three categories of the bus are summarized in Table 1.

Table 1. Categorization of Buses

| No | Bus Types | Bus Variables | | | |
|---|---|---|---|---|---|
| | | $V_k$ | $\delta_k$ | $P_k$ | $Q_k$ |
| 1 | Slack Bus | Known | Known | Not Known | Not known |
| 2 | Generator Bus (PV) | Known | Not known | Known | Not known |
| 3 | Load Bus (PQ) | Not known | Not known | Known | Known |

### B. Power Flow Analysis

The generation of Y-bus admittance using the available data about transformers and transmission lines is the starting point in carrying out the power flow study. Using the Y-bus admittance matrix, the equations for each node in a power system with generators, buses, and loads may be determined as follows [1]–[3]:

$$I = Y_{bus}V \tag{2}$$

For a $n$ bus system, a generalised form of the nodal equation may be expressed as:

$$I_i = \sum_{j=1}^{n} Y_{ij}V_j, for\ i = 1,2,3, \dots n \tag{3}$$

The complex power supplied to bus $i$ is:

$$P_i + jQ_i = V_i I_i^* \tag{4}$$

where,

$$I_i = \frac{P_i - jQ_i}{V_i^*} \tag{5}$$

When $I_i$ is substituted in terms of $P_i$ and $Q_i$, equation (3) becomes:

$$\frac{P_i - jQ_i}{V_i^*} = V_i \sum_{j=1}^{n} Y_{ij} - \sum_{j=1}^{n} Y_{ij}V_j, j \neq i \tag{6}$$

Equation (6) above solves load flow problems using iterative methods, such as NR. Following an analysis of the solution, all relevant data is extracted, which is then used for the real-time configuration of the power system [1], [4]. The following subsection presents the analysis of power systems using the NR method.

i) Power Flow Analysis Based on NR

A Taylor series expansion is used in the NR approach to determine a collection of nonlinear equations from a collection of linear equations [1]. The terms are limited to the first estimate. Since the NR approach can resolve problems that cause divergence with other well-known processes, its convergence features are generally more robust than those of other alternative processes, making it the most iterative approach used for load flow analysis [1]. Also, the admittance matrix is employed when determining equations for currents entering a power system. The polar version of equation (3) shows that *j* contains bus *I* [1]–[3]. Thus,

$$I_i = \sum_{j=1}^{n} |Y_{ij}||V_j| < \theta_{ij} + \delta_j \qquad (7)$$

At bus *i* the real and reactive power is:

$$P_i + jQ_i = V_i^* I_i \qquad (8)$$

Using (8) to replace $I_i$ in (7), we have:

$$P_i - jQ_i = |V_i| < -\delta_i \sum_{j=1}^{n} |Y_{ij}||V_j| < \theta_{ij} + \delta_j \qquad (9)$$

The real and imaginary parts are separated as:

$$P_i = \sum_{j=1}^{n} |V_i||V_j||Y_{ij}| cos(\theta_{ij} - \delta_i + \delta_j) \qquad (10)$$

$$Q_i = -\sum_{j=1}^{n} |V_i||V_j||Y_{ij}| sin(\theta_{ij} - \delta_i + \delta_j) \qquad (11)$$

A collection of non-linear algebraic equations in terms of |V| per unit and δ in radians is represented by (10) and (11). The following set of linear equations is produced by expanding (10) and (11) in Taylor series with respect to the original estimate while ignoring any higher-order terms [1]–[3].

$$\begin{bmatrix} \Delta P_2^{(k)} \\ \vdots \\ \Delta P_n^{(k)} \\ \Delta Q_2^{(k)} \\ \vdots \\ \Delta Q_n^{(k)} \end{bmatrix} = \left[\begin{array}{ccc|ccc} \frac{\partial P_2^{(k)}}{\partial \delta_2} & \cdots & \frac{\partial P_2^{(k)}}{\partial \delta_n} & \frac{\partial P_2^{(k)}}{\partial |V_2|} & \cdots & \frac{\partial P_2^{(k)}}{\partial |V_n|} \\ \vdots & \ddots & \vdots & \vdots & \ddots & \vdots \\ \frac{\partial P_n^{(k)}}{\partial \delta_2} & \cdots & \frac{\partial P_2^{(k)}}{\partial \delta_n} & \frac{\partial P_n^{(k)}}{\partial |V_2|} & \cdots & \frac{\partial P_n^{(k)}}{\partial |V_n|} \\ \hline \frac{\partial Q_2^{(k)}}{\partial \delta_2} & \cdots & \frac{\partial Q_2^{(k)}}{\partial \delta_2} & \frac{\partial Q_2^{(k)}}{\partial |V_2|} & \cdots & \frac{\partial Q_2^{(k)}}{\partial |V_2|} \\ \vdots & \ddots & \vdots & \vdots & \ddots & \vdots \\ \frac{\partial Q_n^{(k)}}{\partial \delta_2} & \cdots & \frac{\partial Q_n^{(k)}}{\partial \delta_n} & \frac{\partial Q_2^{(k)}}{\partial |V_2|} & \cdots & \frac{\partial Q_2^{(k)}}{\partial |V_2|} \end{array}\right] \begin{bmatrix} \Delta \delta_2^{(k)} \\ \vdots \\ \Delta \delta_n^{(k)} \\ \Delta |V_2^{(k)}| \\ \vdots \\ \Delta |V_n^{(k)}| \end{bmatrix} \qquad (12)$$

Since they are previously known, the slack bus (bus 1) variable voltage magnitude and angle are not included in the calculation above. The partial derivatives of (10) and (11), which provide a linearised connection between slight variations in real power, $\Delta P_i^{(k)}$, reactive power, $\Delta Q_i^{(k)}$, voltage magnitude, $\Delta|V_i^{(k)}|$, voltage angle, $\Delta\delta_i^{(k)}$ are expressed to yield the elements of the Jacobian matrix. The following is a matrix representation of equation (12) [1]–[3]:

$$\begin{bmatrix} \Delta P \\ \Delta Q \end{bmatrix} = \begin{bmatrix} J_1 & J_3 \\ J_2 & J_4 \end{bmatrix} \begin{bmatrix} \Delta \delta \\ \Delta |V| \end{bmatrix} \qquad (13)$$

where, $J_1, J_2, J_3$, and $J_4$ are the Jacobian matrix's constituent components.

For the terms $\Delta P_i^{(k)}$ and $\Delta Q_i^{(k)}$, the power residuals—the difference between the scheduled and computed parameters, are determined as follows:

$$\Delta P_i^{(k)} = P_i^{sch} - P_i^{(k)}$$
$$\Delta Q_i^{(k)} = Q_i^{sch} - Q_i^{(k)} \qquad (14)$$

The updated bus voltage estimations are determined as:

$$\delta_i^{(k+1)} = \delta_i^{(k)} + \Delta\delta_i^{(k)} \qquad (15)$$

$$\Delta|V_i^{(k+1)}| = |V_i^{(k)}| + \Delta|V_i^{(k)}| \qquad (16)$$

## III. Design Methodology and System Modelling

The single-line diagram of the 5-bus power system considered in this study is shown in Fig. 2. The system comprises two generators, two transformers, transmission lines, circuit breakers, and 5 buses. Also, bus 1 is considered the slack bus. Furthermore, $S_{base}$ = 100 MVA, $V_{base}$ = 15 kV at buses 1 and 3, and 345 kV at buses 2, 4, and 5. The parameters of the system are presented in Tables 2 – 4 [3].

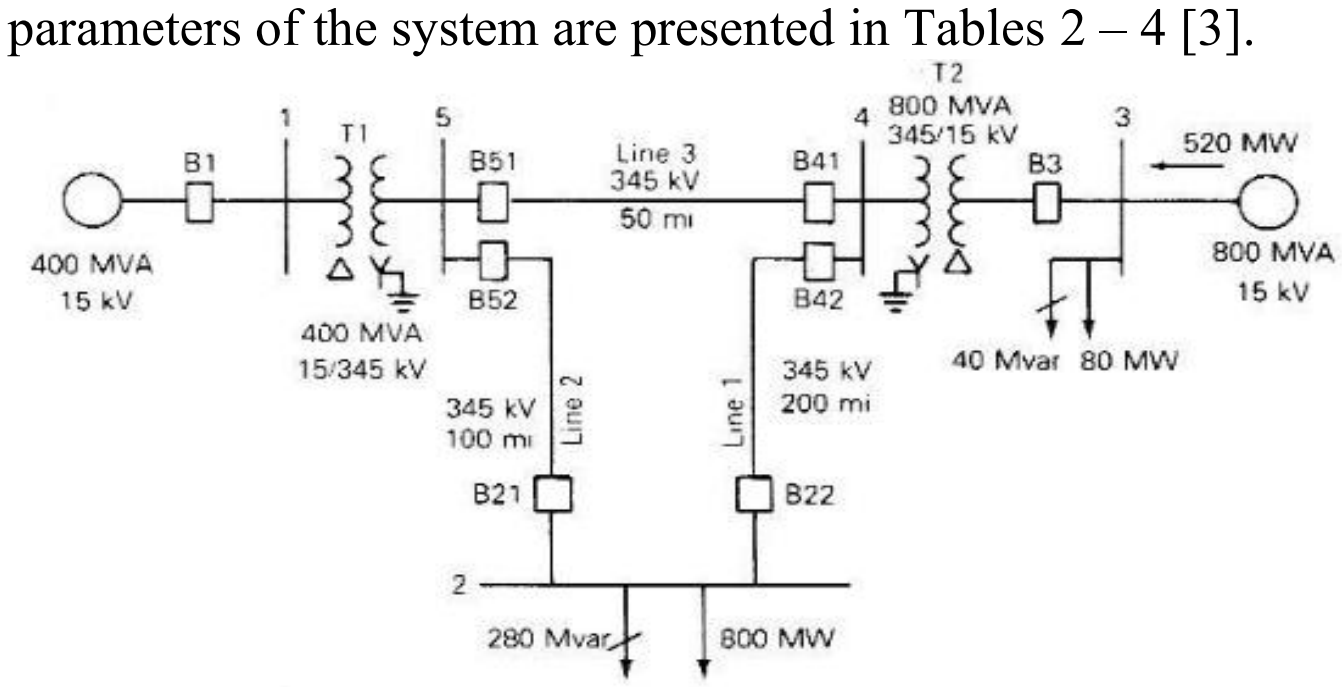

Fig. 2. Single-line diagram of a 5-bus power system [3]

Table 2. Bus parameters

| Bus | Bus Types | Bus Variables | | | | | | | |
|---|---|---|---|---|---|---|---|---|---|
| | | $V$ pu | $\delta$ deg | $P_G$ pu | $Q_G$ pu | $P_L$ pu | $Q_L$ pu | $Q_{Gmax}$ pu | $Q_{Gmin}$ pu |
| 1 | Slack | 1.0 | 0 | - | - | 0 | 0 | - | - |
| 2 | Load | - | - | 0 | 0 | 8.0 | 2.8 | - | - |
| 3 | Constant Voltage | 1.0 5 | - | 5.2 | - | 0.8 | 0.4 | 4.0 | -2.8 |
| 4 | Load | - | - | 0 | 0 | 0 | 0 | - | - |
| 5 | Load | - | - | 0 | 0 | 0 | 0 | - | - |

Table 3. Bus parameters

| Bus-to-bus | Bus Variables | | | | |
|---|---|---|---|---|---|
| | $R'$ pu | $X'$ pu. | $G'$ pu | $B'$ pu | Maximum MVA pu |
| 2-4 | 0.0090 | 0.100 | 0 | 1.72 | 12.0 |
| 2-5 | 0.0045 | 0.050 | 0 | 0.88 | 12.0 |
| 4-5 | 0.00225 | 0.025 | 0 | 0.44 | 12.0 |

Table 4. Bus parameters

| Bus-to-bus | Bus Variables | | | | | |
|---|---|---|---|---|---|---|
| | $R$ pu | $X$ pu. | $G_c$ pu | $B_m$ pu | Maximum MVA pu | Maximum TAP Setting pu |
| 1-5 | 0.00150 | 0.02 | 0 | 0 | 6.0 | - |
| 3-4 | 0.00075 | 0.01 | 0 | 0 | 10.0 | - |

Using the parameters above, the system (base case) was modeled in PowerWorld Simulator (PWS), as shown in Fig. 2. Here, the transformer taps are set to 1.0. The system configurations on the PWS are shown in Fig. 3 – 4, whereas the Ybus matrix determined using PWS is shown in Fig. 5.

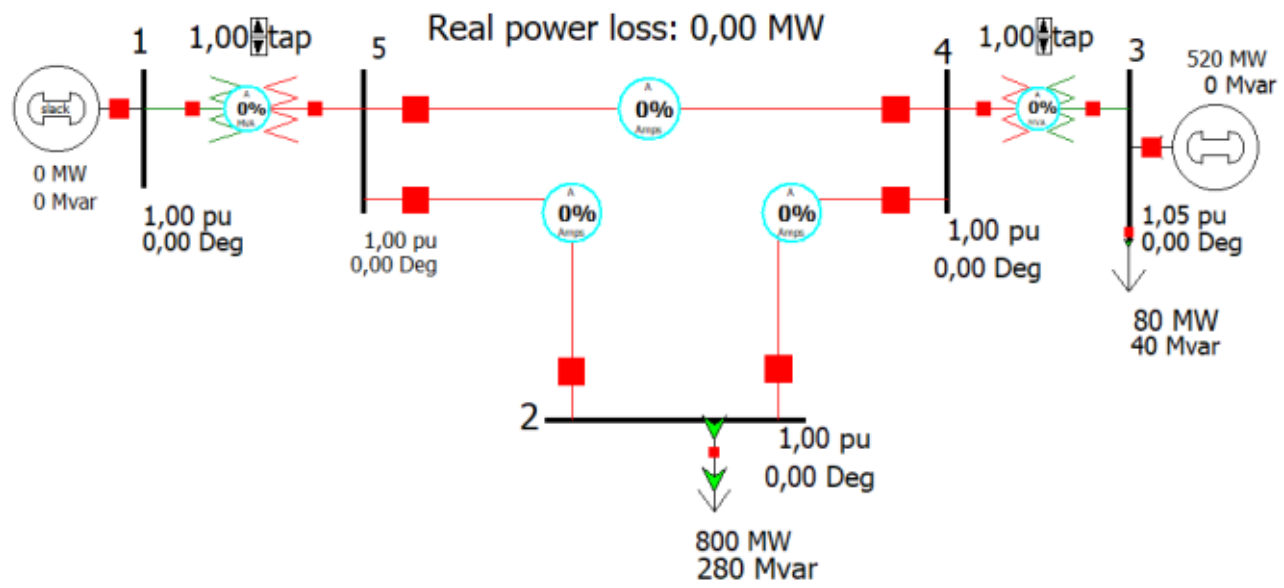

Fig. 2. Modelled 5-bus power system in PWS

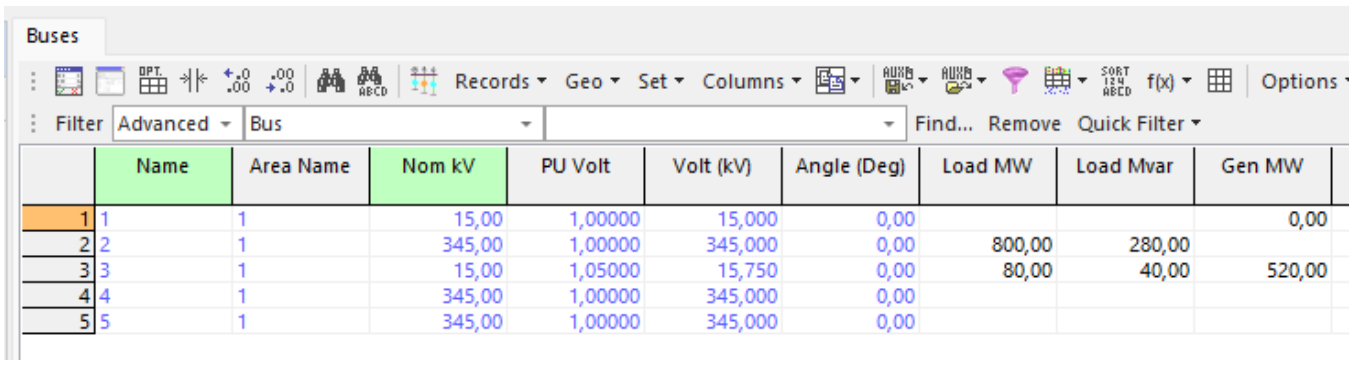

| | Number | Name | Area Name | Nom kV | PU Volt | Volt (kV) | Angle (Deg) | Load MW | Load Mvar | Gen MW |
|---|---|---|---|---|---|---|---|---|---|---|
| 1 | 1 | 1 | 1 | 15,00 | 1,00000 | 15,000 | 0,00 | | | 0,00 |
| 2 | 2 | 2 | 1 | 345,00 | 1,00000 | 345,000 | 0,00 | 800,00 | 280,00 | |
| 3 | 3 | 3 | 1 | 15,00 | 1,05000 | 15,750 | 0,00 | 80,00 | 40,00 | 520,00 |
| 4 | 4 | 4 | 1 | 345,00 | 1,00000 | 345,000 | 0,00 | | | |
| 5 | 5 | 5 | 1 | 345,00 | 1,00000 | 345,000 | 0,00 | | | |

Fig. 3. System buses parameters

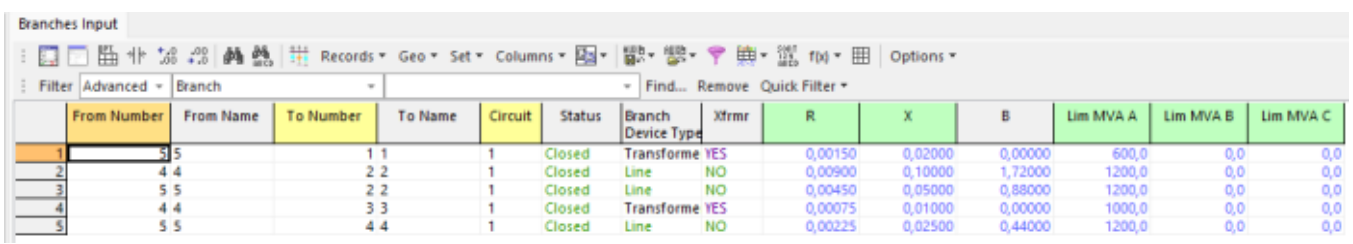

| | From Number | From Name | To Number | To Name | Circuit | Status | Branch Device Type | Xfrmr | R | X | B | Lim MVA A | Lim MVA B | Lim MVA C |
|---|---|---|---|---|---|---|---|---|---|---|---|---|---|---|
| 1 | 5 | 5 | 1 | 1 | 1 | Closed | Transforme | YES | 0,00150 | 0,02000 | 0,00000 | 600,0 | 0,0 | 0,0 |
| 2 | 4 | 4 | 2 | 2 | 1 | Closed | Line | NO | 0,00900 | 0,10000 | 1,72000 | 1200,0 | 0,0 | 0,0 |
| 3 | 5 | 5 | 2 | 2 | 1 | Closed | Line | NO | 0,00450 | 0,05000 | 0,88000 | 1200,0 | 0,0 | 0,0 |
| 4 | 4 | 4 | 3 | 3 | 1 | Closed | Transforme | YES | 0,00075 | 0,01000 | 0,00000 | 1000,0 | 0,0 | 0,0 |
| 5 | 5 | 5 | 4 | 4 | 1 | Closed | Line | NO | 0,00225 | 0,02500 | 0,44000 | 1200,0 | 0,0 | 0,0 |

Fig. 4. System branches input parameters

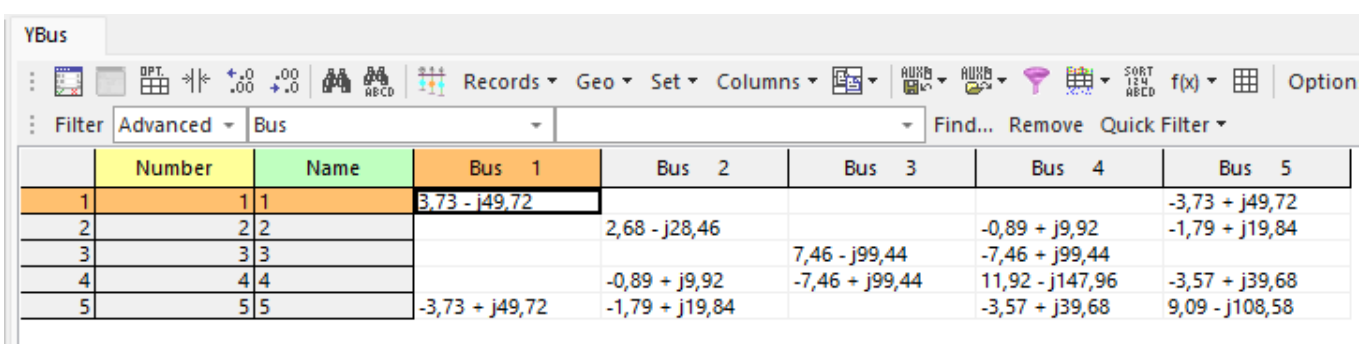

| | Number | Name | Bus 1 | Bus 2 | Bus 3 | Bus 4 | Bus 5 |
|---|---|---|---|---|---|---|---|
| 1 | 1 | 1 | 3,73 - j49,72 | | | | -3,73 + j49,72 |
| 2 | 2 | 2 | | 2,68 - j28,46 | | -0,89 + j9,92 | -1,79 + j19,84 |
| 3 | 3 | 3 | | | 7,46 - j99,44 | -7,46 + j99,44 | |
| 4 | 4 | 4 | | -0,89 + j9,92 | -7,46 + j99,44 | 11,92 - j147,96 | -3,57 + j39,68 |
| 5 | 5 | 5 | -3,73 + j49,72 | -1,79 + j19,84 | | -3,57 + j39,68 | 9,09 - j108,58 |

Fig. 5. System Ybus matrix parameters

## IV. Simulation Results and Discussions

### *A. Base Case System Simulation Results Based on PWS*

In this section, the system, referred to as the base case system, was simulated for 50 iterations. Also, the transformer taps are set to 1.0. The simulated system model is shown in Fig. 6, whereas the system buses and branches parameters are shown in Fig. 7 and Fig. 8, respectively. In this study, the normal range of the bus voltages is assumed to be from 0.95 per unit (pu) to 1.5 pu. As shown in Fig. 6 and Fig. 7, all the buses satisfied this requirement, except for bus 2, which has 0.83 pu. Also, the only parts of the system with loading of 50% or more are the two transformers and the transmission line between buses 2 and 5, whereas other transmission lines experienced loading of less than 30%. Furthermore, the system experienced a total real power loss of 34.84 MW.

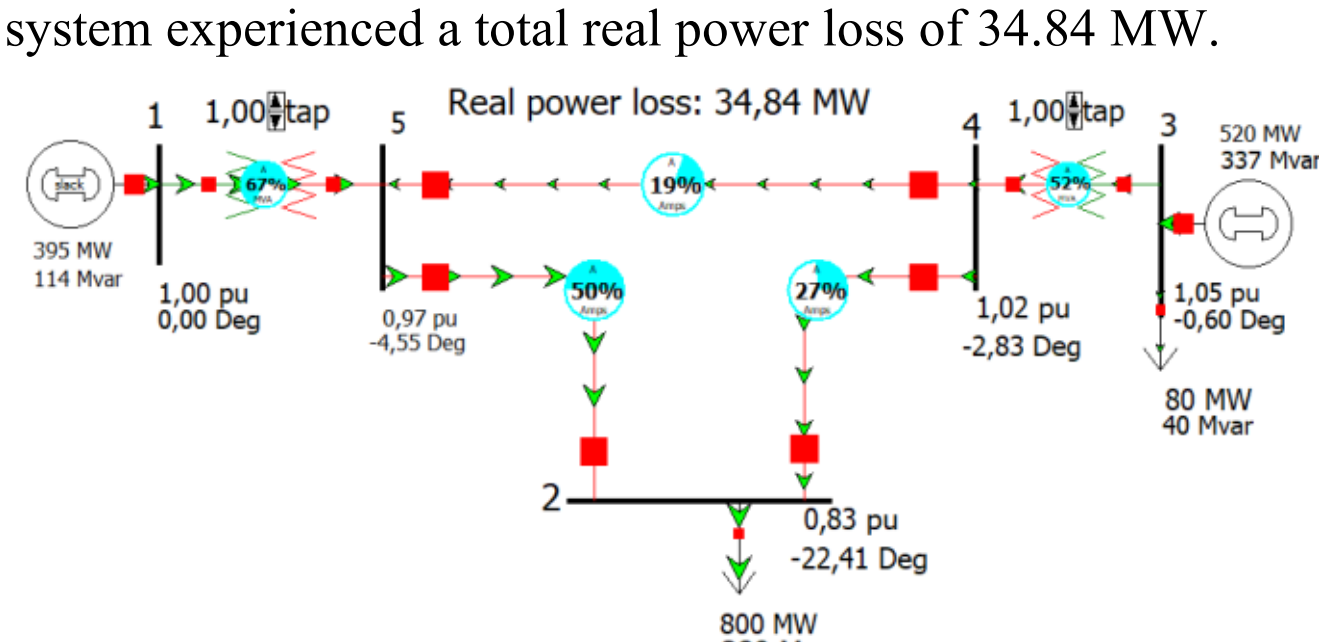

Fig. 6. Simulated 5-bus power system (base case) in PWS

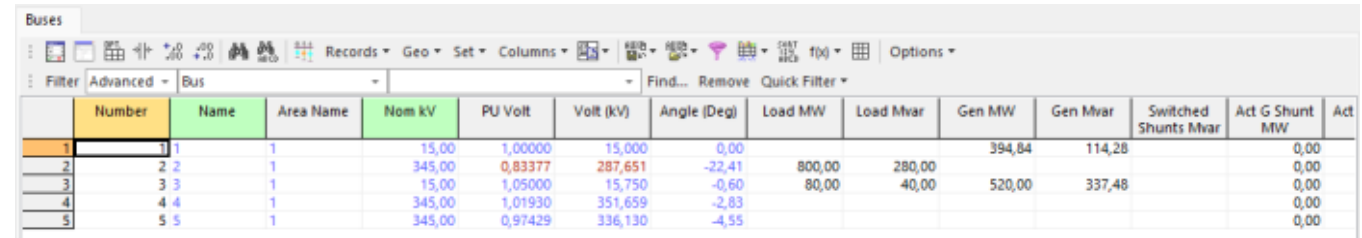

| | Number | Name | Area Name | Nom kV | PU Volt | Volt (kV) | Angle (Deg) | Load MW | Load Mvar | Gen MW | Gen Mvar | Switched Shunts Mvar | Act G Shunt MW |
|---|---|---|---|---|---|---|---|---|---|---|---|---|---|
| 1 | 1 | 1 | 1 | 15,00 | 1,00000 | 15,000 | 0,00 | | | 394,84 | 114,28 | | 0,00 |
| 2 | 2 | 2 | 1 | 345,00 | 0,83377 | 287,651 | -22,41 | 800,00 | 280,00 | | | | 0,00 |
| 3 | 3 | 3 | 1 | 15,00 | 1,05000 | 15,750 | -0,60 | 80,00 | 40,00 | 520,00 | 337,48 | | 0,00 |
| 4 | 4 | 4 | 1 | 345,00 | 1,01930 | 351,659 | -2,83 | | | | | | 0,00 |
| 5 | 5 | 5 | 1 | 345,00 | 0,97429 | 336,130 | -4,55 | | | | | | 0,00 |

Fig. 7. System bus parameters

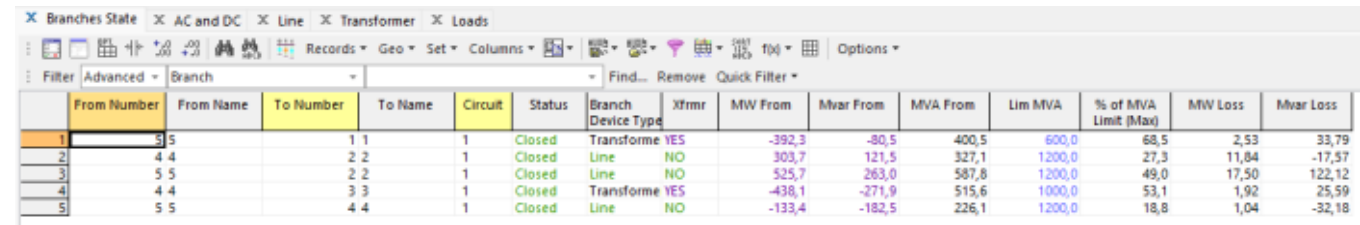

| | From Number | From Name | To Number | To Name | Circuit | Status | Branch Device Type | Xfrmr | MW From | Mvar From | MVA From | Lim MVA | % of MVA Limit (Max) | MW Loss | Mvar Loss |
|---|---|---|---|---|---|---|---|---|---|---|---|---|---|---|---|
| 1 | 5 | 5 | 1 | 1 | 1 | Closed | Transforme | YES | -392,3 | -80,5 | 400,5 | 600,0 | 68,5 | 2,53 | 33,79 |
| 2 | 4 | 4 | 2 | 2 | 1 | Closed | Line | NO | 303,7 | 121,5 | 327,1 | 1200,0 | 27,3 | 11,84 | -17,57 |
| 3 | 5 | 5 | 2 | 2 | 1 | Closed | Line | NO | 525,7 | 263,0 | 587,8 | 1200,0 | 49,0 | 17,50 | 122,12 |
| 4 | 4 | 4 | 3 | 3 | 1 | Closed | Transforme | YES | -438,1 | -271,9 | 515,6 | 1000,0 | 53,1 | 1,92 | 25,59 |
| 5 | 5 | 5 | 4 | 4 | 1 | Closed | Line | NO | -133,4 | -182,5 | 226,1 | 1200,0 | 18,8 | 1,04 | -32,18 |

Fig. 8. System branches state

### *B. Base Case System Simulation Results Based on MATLAB Software*

A MATLAB code was developed to solve the power flow problem and simulated for the same number of iterations as the PWS method. The simulation results are presented in Table 5. As shown in the table, the voltage values of the system buses correspond to the values obtained using PWS software. However, there is a huge discrepancy in the values of the voltage angles and real and reactive power obtained. Improving the results may require higher scaling of the simulation parameter or a higher number of iterations. This shows that the PWS is more suitable for performing power flow analysis.

Table 5. Power flow study results based on MATLAB simulation

| Bus | Bus Types | Variables | | | |
|---|---|---|---|---|---|
| | | *V* pu | *δ* deg | *P* (MW) | *Q* (MVar) |
| 1 | Slack | 1.0 | 0 | 3.95 | 1.14 |
| 2 | Load | 0.83377 | 0.391 | -8.0 | -2.80 |
| 3 | Constant Voltage | 1.05 | 0.0104 | 4.40 | 2.98 |
| 4 | Load | 1.019 | -0.049 | -0.000007 | 0.0000006 |
| 5 | Load | 0.974 | -0.079 | -0.000014 | 0.000002 |

### *C. Voltage Control of Bus 2*

Static VAR systems, shunt reactors, and shunt capacitor banks are the systems often used to control the system power flows. In addition, transformer regulation and tap-changing control are other techniques employed [3]. In this study, the voltage control at bus 2, using a switched shunt capacitor bank and tap changing control, is evaluated.

#### *1) Voltage Control Using a Shunt Capacitor Bank*

As shown in Fig. 9 – 11, when a reactive shunt capacitor rated 190-Mvar (nominal value) is added to bus 2 with the load (800MW + j280 MVar), the voltage level of the bus increases to 0.95 pu, from 0.83 pu. The switched shunt capacitor mainly mitigates the capacitive impact of the transmission lines, including power and current loss reduction. Thus, the total real power loss is reduced to 25.68 MW. In addition, the loading on the lines was reduced by more than 2%. Fig. 9 further shows that the shunt capacitor only delivered 172.4 Mvar of reactive power. This disparity occurs because the squared voltage terminal $\left(Q_c = \frac{V_C^2}{X_c}\right)$ affects the capacitor's reactive output. Here, an assumed voltage of 1.0 pu is used to get a capacitor's Mvar rating [3].

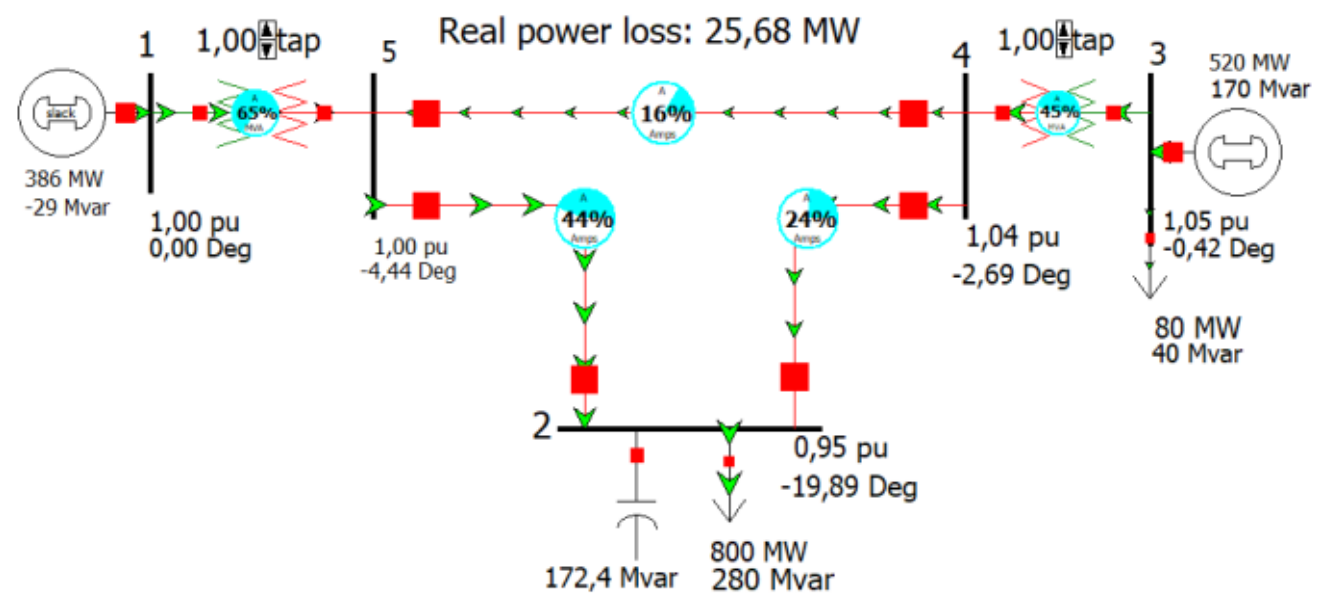

Fig. 9. Simulated 5-bus power system with a shunt capacitor bank

| | Number | Name | Area Name | Nom kV | PU Volt | Volt (kV) | Angle (Deg) | Load MW | Load Mvar | Gen MW | Gen Mvar | Switched Shunts Mvar | Act G Shunt MW | Act B Shunt Mvar |
|---|---|---|---|---|---|---|---|---|---|---|---|---|---|---|
| 1 | 1 | 1 | 1 | 15,00 | 1,00000 | 15,000 | 0,00 | | | 385,67 | -28,68 | | 0,00 | 0,00 |
| 2 | 2 | 2 | 1 | 345,00 | 0,95254 | 328,628 | -19,89 | 800,00 | 280,00 | | | 172,39 | 0,00 | 0,00 |
| 3 | 3 | 3 | 1 | 15,00 | 1,05000 | 15,750 | -0,42 | 80,00 | 40,00 | 520,00 | 169,64 | | 0,00 | 0,00 |
| 4 | 4 | 4 | 1 | 345,00 | 1,03532 | 357,186 | -2,69 | | | | | | 0,00 | 0,00 |
| 5 | 5 | 5 | 1 | 345,00 | 1,00295 | 346,019 | -4,44 | | | | | | 0,00 | 0,00 |

Fig. 10. System Buses parameters

| | From Number | From Name | To Number | To Name | Circuit | Status | Branch Device Type | Xfrmr | MW From | Mvar From | MVA From | Lim MVA | % of MVA Limit (Max) | MW Loss | Mvar Loss |
|---|---|---|---|---|---|---|---|---|---|---|---|---|---|---|---|
| 1 | 5 | 5 | 1 | 1 | 1 | Closed | Transforme | YES | -383,4 | 58,6 | 387,9 | 600,0 | 64,6 | 2,24 | 29,91 |
| 2 | 4 | 4 | 2 | 2 | 1 | Closed | Line | NO | 300,9 | 10,6 | 301,1 | 1200,0 | 25,4 | 8,49 | -75,89 |
| 3 | 5 | 5 | 2 | 2 | 1 | Closed | Line | NO | 520,4 | 79,1 | 526,3 | 1200,0 | 43,9 | 12,79 | 57,98 |
| 4 | 4 | 4 | 3 | 3 | 1 | Closed | Transforme | YES | -438,6 | -110,6 | 452,3 | 1000,0 | 45,9 | 1,43 | 19,08 |
| 5 | 5 | 5 | 4 | 4 | 1 | Closed | Line | NO | -136,9 | -137,7 | 194,2 | 1200,0 | 16,2 | 0,72 | -37,73 |

Fig. 11. System branches state

## D. Changing Taps of Transformers

A series of admittance, $y_t$ in pu is used to describe a tap-changing transformer when the ratio is at the nominal value. The pu admittance on both sides of the transformer is different when the off-nominal ratio is present, and the admittance needs to be adjusted to account for the off-nominal ratio's influence [2]. Suppose that an ideal transformer that represents the off-nominal tap ratio $1{:}\alpha$ is connected in series with a transformer that has admittance. $y_t$ is the pu off-nominal tap location that permits a minor voltage modification of typically +10%, and it is the admittance in pu based on the nominal turn ratio. $\alpha$ is a complex value in the context of phase-changing transformers [2]. This concept is verified in this study by varying the transformer taps from 0.85 to 1.15 in increments of 0.05. The goal is to increase the voltage level at bus 2 to 0.95 pu, from 0.83. The control settings in PWS are shown in Fig. 12.

| | Circuit | Type | Status | Tap Ratio | Phase (Deg) | Automatic Control | Reg Bus Num | Reg Value | Reg Error | Reg Min | Reg Max | Tap Min | Tap Max | Step Size |
|---|---|---|---|---|---|---|---|---|---|---|---|---|---|---|
| 1 | 1 | LTC | Closed | 1,00000 | 0,00000 | NO | 5 | 0,97429 | -0,01571 | 0,99000 | 1,00000 | 0,85000 | 1,15000 | 0,05000 |
| 2 | 1 | LTC | Closed | 1,00000 | 0,00000 | NO | 4 | 1,01930 | 0,01930 | 0,99000 | 1,00000 | 0,85000 | 1,15000 | 0,05000 |

Fig. 12. Transformer taps changing control settings

As shown in Fig. 13, setting the tap range of the transformers to 1.06 and 1.07 helped to increase the voltage level at bus 2 to 0.95 pu. Also, Fig. 13 and Fig. 14 show that the real power loss is reduced to 27.52 MW, from 34.84 MW obtained in the base case.

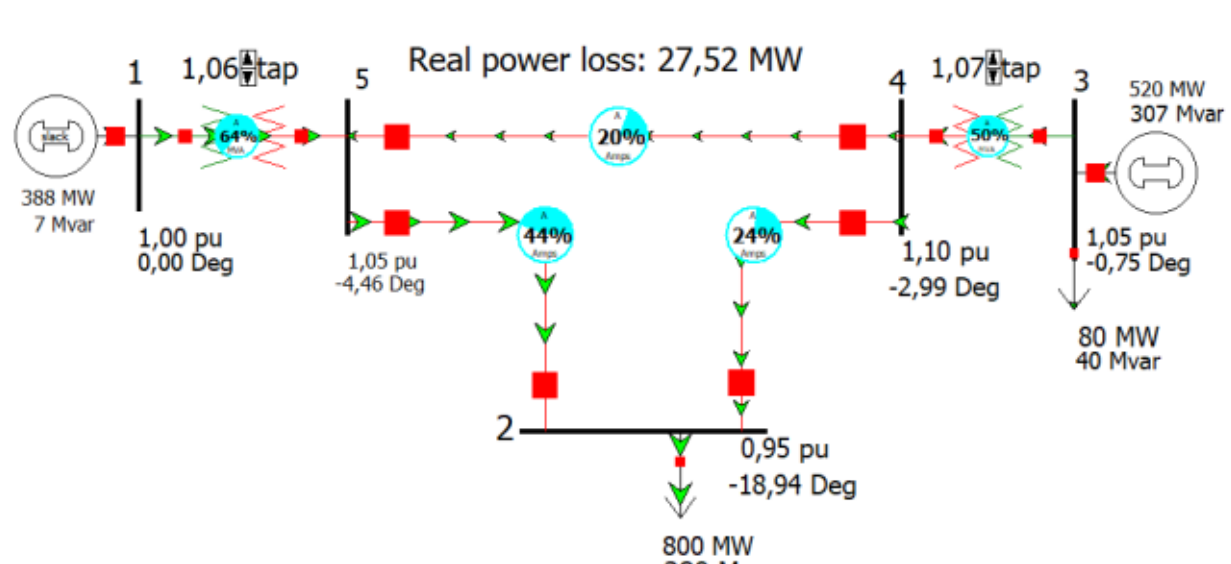

Fig. 13. Simulated 5-bus power system with transformer tap settings

| | From Number | From Name | To Number | To Name | Circuit | Status | Branch Device Type | Xfrmr | MW From | Mvar From | MVA From | Lim MVA | % of MVA Limit (Max) | MW Loss | Mvar Loss |
|---|---|---|---|---|---|---|---|---|---|---|---|---|---|---|---|
| 1 | 5 | 5 | 1 | 1 | 1 | Closed | Transforme | YES | -385,2 | 23,2 | 385,9 | 600,0 | 64,6 | 2,25 | 30,04 |
| 2 | 4 | 4 | 2 | 2 | 1 | Closed | Line | NO | 302,4 | 74,2 | 311,3 | 1200,0 | 27,6 | 9,18 | -79,25 |
| 3 | 5 | 5 | 2 | 2 | 1 | Closed | Line | NO | 520,0 | 185,2 | 552,0 | 1200,0 | 46,0 | 13,22 | 58,66 |
| 4 | 4 | 4 | 3 | 3 | 1 | Closed | Transforme | YES | -438,2 | -243,4 | 501,3 | 1000,0 | 51,5 | 1,80 | 24,05 |
| 5 | 5 | 5 | 4 | 4 | 1 | Closed | Line | NO | -134,8 | -208,4 | 248,2 | 1200,0 | 20,7 | 1,06 | -39,15 |

Fig. 14. System branches state

## E. Adding a New Transformer Between Buses 1 and 5

In this section, an additional transformer is connected in parallel with the existing transformer in buses 1 and 5 in the base case. The parameters of the transformer are similar to the existing one. The goal is to reduce the high loading between buses 1 and 5 observed in the base case. The transformer tap settings are set to 1.0. As shown in Fig. 15 and Fig. 16, adding a new transformer between buses 1 and 5 balanced the loading between the two buses. This reduced the MVA, reactive power, and real power delivered from bus 1 to 5, hence reducing thermal stress and improving reliability, compared to the values obtained in the base case (see Fig. 8). However, in practice, adding the new transformer could potentially lead to an increase in the overall cost of the system.

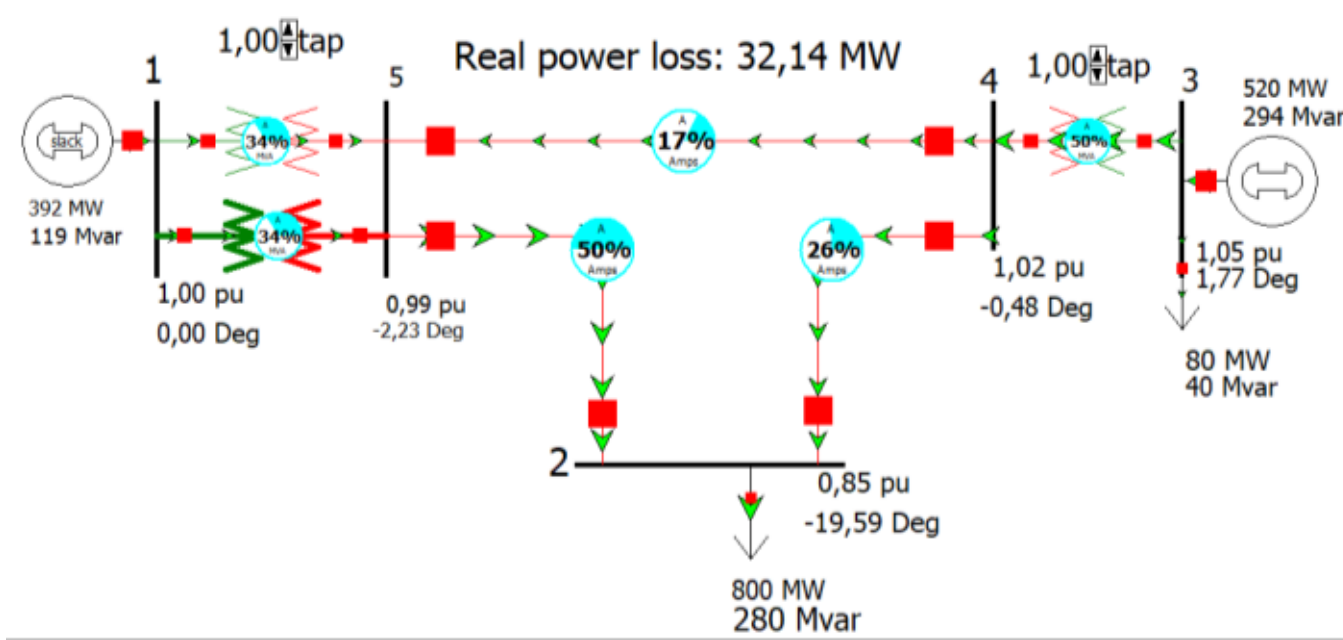

Fig. 15. Simulated 5-bus system with a new transformer connected in parallel between buses 1 and 5.

| | From Number | From Name | To Number | To Name | Circuit | Status | Branch Device Type | Xfrmr | MW From | Mvar From | MVA From | Lim MVA | % of MVA Limit (Max) | MW Loss | Mvar Loss |
|---|---|---|---|---|---|---|---|---|---|---|---|---|---|---|---|
| 1 | 1 | 1 | 5 | 5 | 2 | Closed | Transforme | YES | 196,1 | 59,6 | 204,9 | 600,0 | 34,2 | 0,63 | 8,40 |
| 2 | 5 | 5 | 1 | 1 | 1 | Closed | Transforme | YES | -195,4 | -51,2 | 202,0 | 600,0 | 34,2 | 0,63 | 8,40 |
| 3 | 4 | 4 | 2 | 2 | 1 | Closed | Line | NO | 302,2 | 110,1 | 321,6 | 1200,0 | 26,8 | 11,29 | -26,50 |
| 4 | 5 | 5 | 2 | 2 | 1 | Closed | Line | NO | 526,1 | 257,9 | 585,9 | 1200,0 | 48,8 | 17,00 | 114,46 |
| 5 | 4 | 4 | 3 | 3 | 1 | Closed | Transforme | YES | -438,2 | -230,5 | 495,2 | 1000,0 | 50,8 | 1,76 | 23,41 |
| 6 | 5 | 5 | 4 | 4 | 1 | Closed | Line | NO | -135,2 | -155,5 | 206,1 | 1200,0 | 17,2 | 0,84 | -35,10 |

Fig. 16. System branches state

## F. Adding a Transmission Line Between Buses 2 and 4

In this section, an additional transmission line is connected in parallel with the existing transformer in buses 2 and 4 in the base case. The parameters of the transmission line are similar to the existing one. The goal is to reduce the loading between buses 2 and 4 and to improve the system capacity. As shown in Fig. 17 and Fig. 18, adding a new transmission line between buses 2 and 4 also balanced the loading between the two lines, significantly reducing the loading between other buses, except for the loading between buses 1 and 5. Compared to the base case, it can be seen in Fig. 17 that the voltage level at bus 2 increased to 0.96 pu. Also, the voltage level at other buses increased without violating the desired voltage requirement. Furthermore, the real power loss is reduced to 19.62 MW, from 34.84 MW obtained in the base case. Again, the major limitation of this approach is that, in practice, adding the new transmission line will lead to an increase in the overall cost of the system.

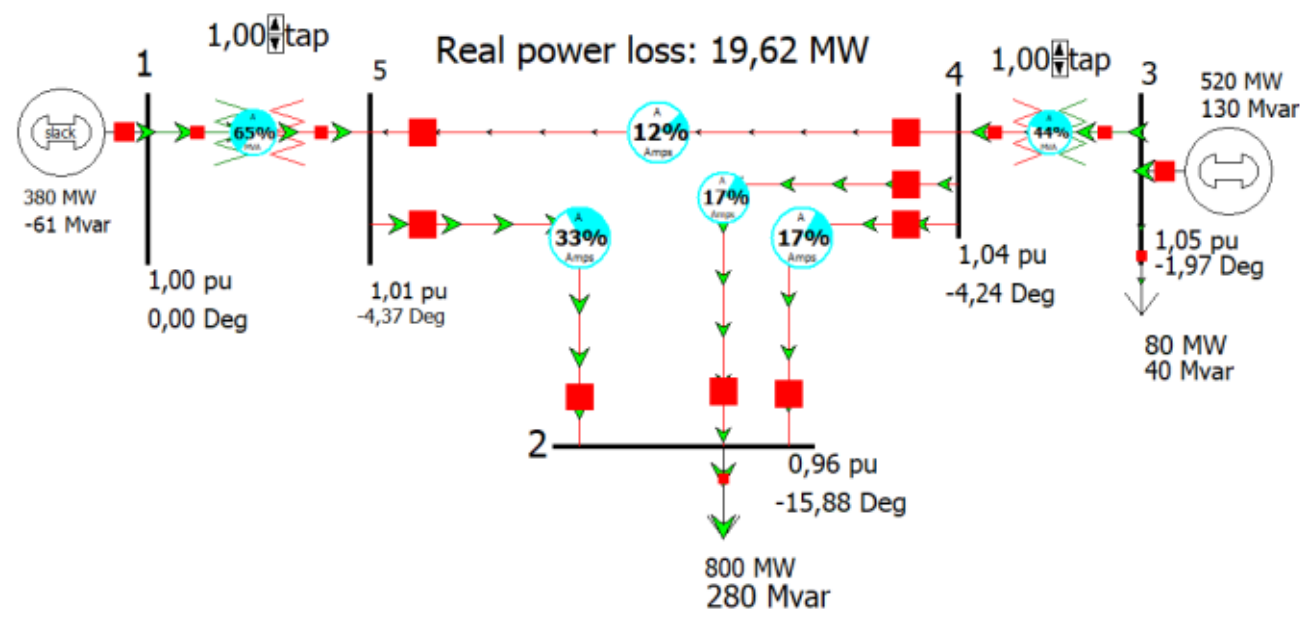

Fig. 17. Simulated 5-bus power system with another transmission line between buses 2 and 4

Branches State

| | From Number | From Name | To Number | To Name | Circuit | Status | Branch Device Type | Xfrmr | MW From | Mvar From | MVA From | Lim MVA | % of MVA Limit (Max) | MW Loss | Mvar Loss |
|---|---|---|---|---|---|---|---|---|---|---|---|---|---|---|---|
| 1 | 5 | 5 | 1 | 1 | 1 | Closed | Transforme | YES | -377,4 | 90,7 | 388,1 | 600,0 | 64,7 | 2,22 | 29,57 |
| 2 | 4 | 4 | 2 | 2 | 2 | Closed | Line | NO | 209,3 | -13,3 | 209,7 | 1200,0 | 19,5 | 4,18 | -126,40 |
| 3 | 4 | 4 | 2 | 2 | 1 | Closed | Line | NO | 209,3 | -13,3 | 209,7 | 1200,0 | 19,5 | 4,18 | -126,40 |
| 4 | 5 | 5 | 2 | 2 | 1 | Closed | Line | NO | 397,1 | 49,8 | 400,2 | 1200,0 | 33,4 | 7,36 | -3,97 |
| 5 | 4 | 4 | 3 | 3 | 1 | Closed | Transforme | YES | -438,6 | -71,3 | 444,4 | 1000,0 | 44,9 | 1,37 | 18,29 |
| 6 | 5 | 5 | 4 | 4 | 1 | Closed | Line | NO | -19,7 | -140,5 | 141,9 | 1200,0 | 11,8 | 0,32 | -42,66 |

Fig. 18. System branches state

### *G. Removing the Transmission Line Between Buses 2 and 5*

In this section, the transmission line between buses 2 and 5 is removed to simulate a maintenance scenario. As shown in Fig. 19a, the initial simulation resulted in a system blackout, causing a severe reduction in the load served at bus 2, dropping from 800 MW and 280 MVar to 184 MW and 65 MVar, respectively. Additionally, the real power loss increased significantly to 73.78 MW, which is more than twice the loss recorded in the base case – an unacceptable outcome. Following the blackout, a second simulation was conducted. As shown in Fig. 19b, although the system remained operational this time, it was only able to supply approximately 23% of the total load demand. Furthermore, the loading on the transmission lines between buses 2 and 4, and buses 3 and 4, increased significantly by more than 26%. While the voltage levels at buses 1, 3, 4, and 5 remained within acceptable limits, the voltage at bus 2 dropped to 0.22 pu, which is far below the required threshold for stable operation.

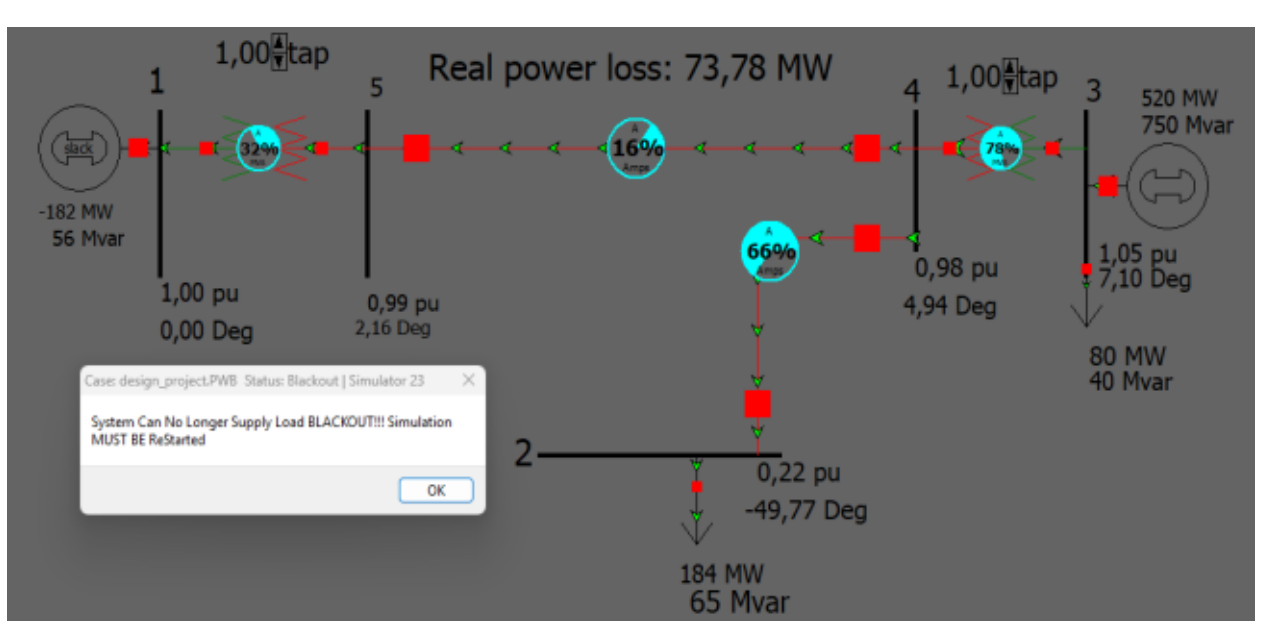

(a)

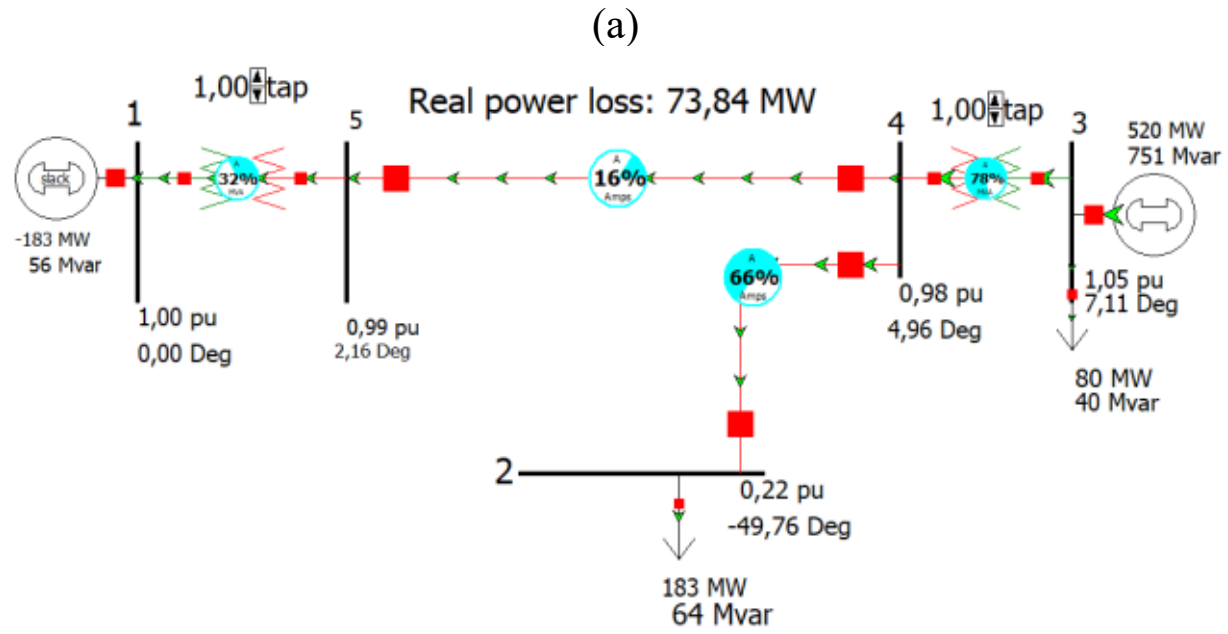

(b)

Fig. 19. Simulated 5-bus power system without a transmission line between buses 2 and 5. (a) First simulation, and (b) Second simulation

To improve the system's performance, a reactive shunt capacitor with a nominal rating of 190 Mvar—similar to the one previously used—was added to bus 2, as illustrated in Fig. 20. Despite this addition, the voltage level at bus 2 only increased marginally to 0.23 pu, which remains significantly below the acceptable threshold for reliable system operation. This result indicates that the shunt compensation alone is insufficient to restore voltage stability under the current operating conditions.

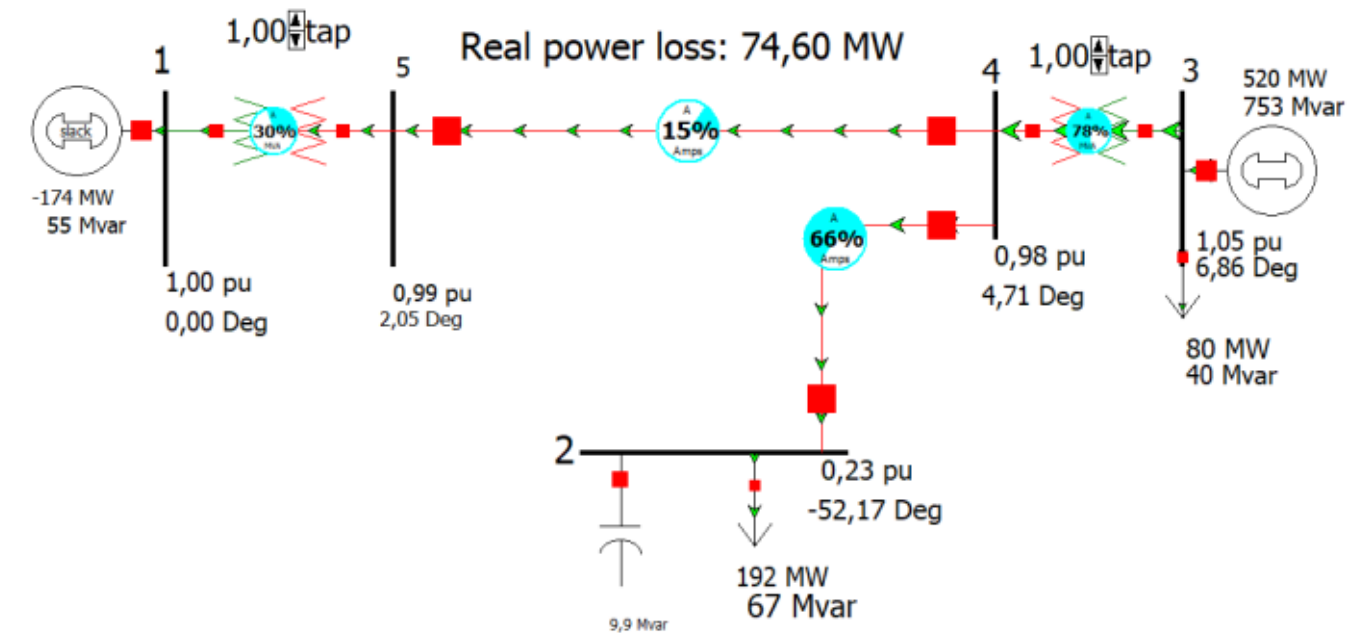

Fig. 20. Simulated 5-bus power system without a transmission line between buses 2 and 5 and with a 190-Mvar shunt capacitor at bus 2.

Increasing the nominal rating of the shunt capacitor to 2000 Mvar resulted in only a modest voltage rise at Bus 2, reaching 0.26 pu, as shown in Fig. 21. However, this reactive power injection led to a noticeable increase in the loading of the transformer between Buses 3 and 4. This result indicates that while the capacitor provided limited voltage support, it simultaneously imposed additional stress on other components of the system.

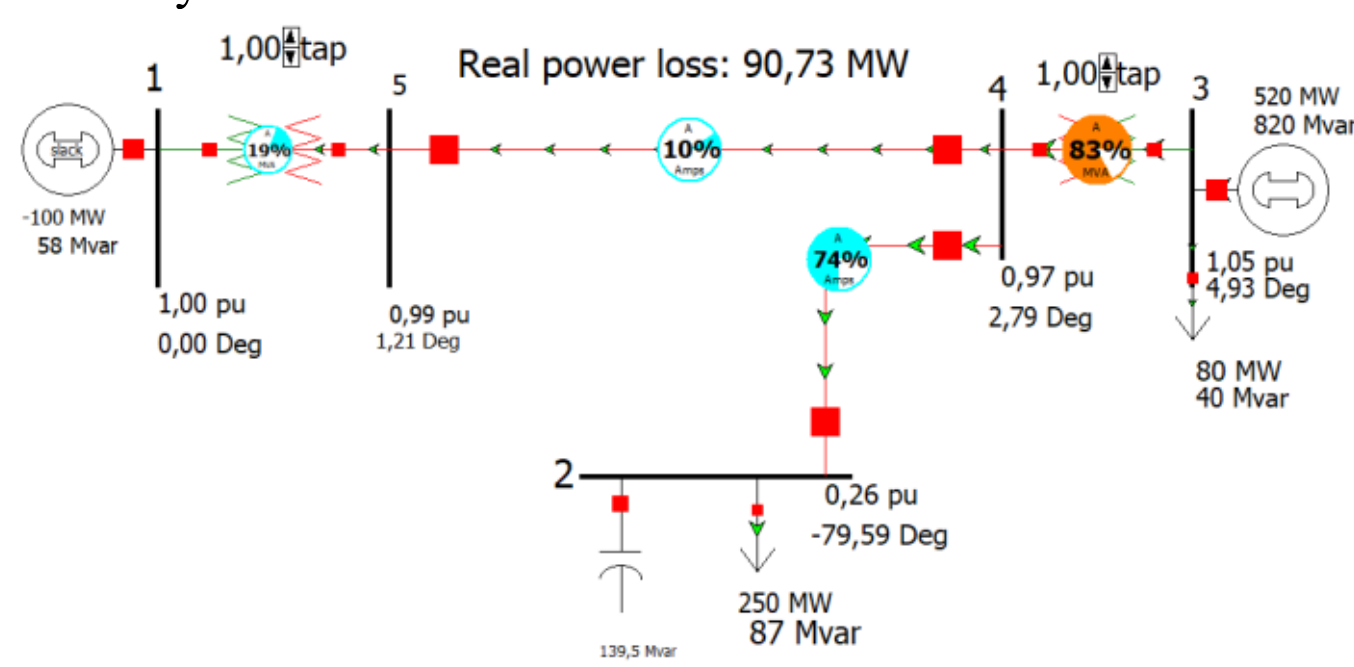

Fig. 21. Simulated 5-bus power system without a transmission line between buses 2 and 5 and with a 2000-Mvar shunt capacitor at bus 2.

To further investigate the voltage behavior, the shunt capacitor capacity at Bus 2 was increased to 4000 Mvar. However, as shown in Fig. 22, this adjustment did not result in a voltage improvement at Bus 2. Instead, the voltage level decreased significantly to 0.23 pu. Additionally, the power loading between Buses 2 and 4, as well as between Buses 4 and 3, increased further. These observations indicate that simply increasing the capacitor size does not improve the voltage profile to within acceptable limits. Therefore, alternative voltage support or network configuration strategies may be considered.

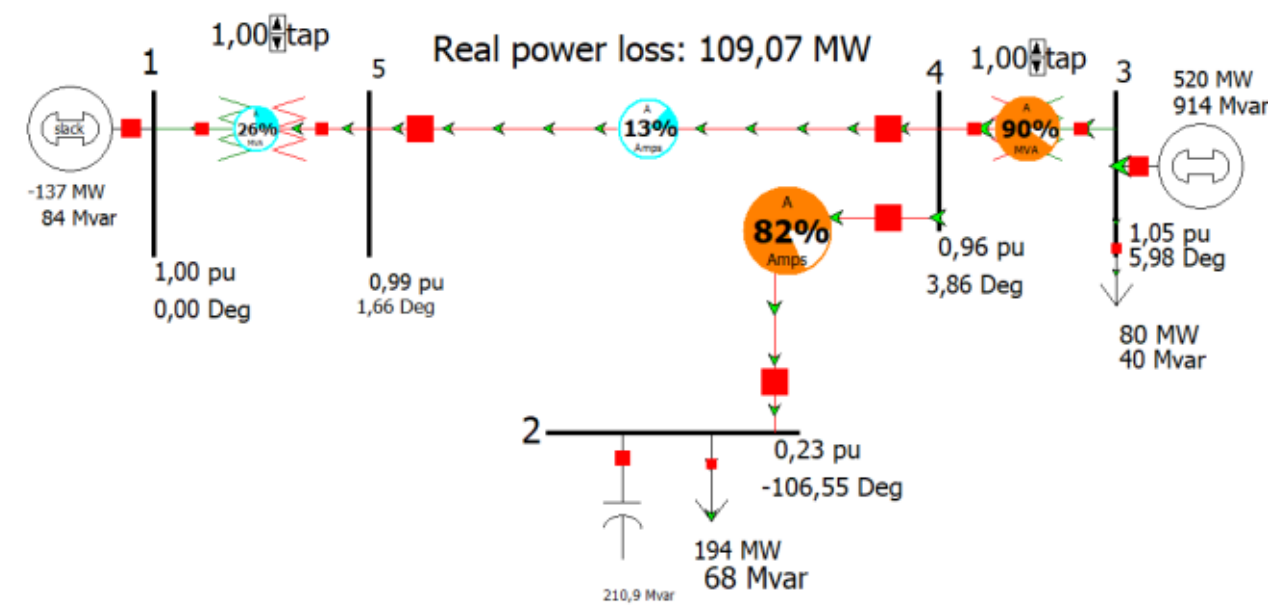

Fig. 22. Simulated 5-bus power system without a transmission line between buses 2 and 5 and with a 4000-Mvar shunt capacitor at bus 2.

## *H. Loadshedding at Bus 2*

In this section, the shunt capacitor at Bus 2 was removed to evaluate the amount of load shedding required to maintain the voltage level above 0.95 pu. As shown in Fig. 23, reducing the load at Bus 2 by up to 35% is sufficient to sustain the voltage at or above 0.95 pu. This indicates that selective load shedding can be an effective strategy for voltage regulation in the absence of reactive power support.

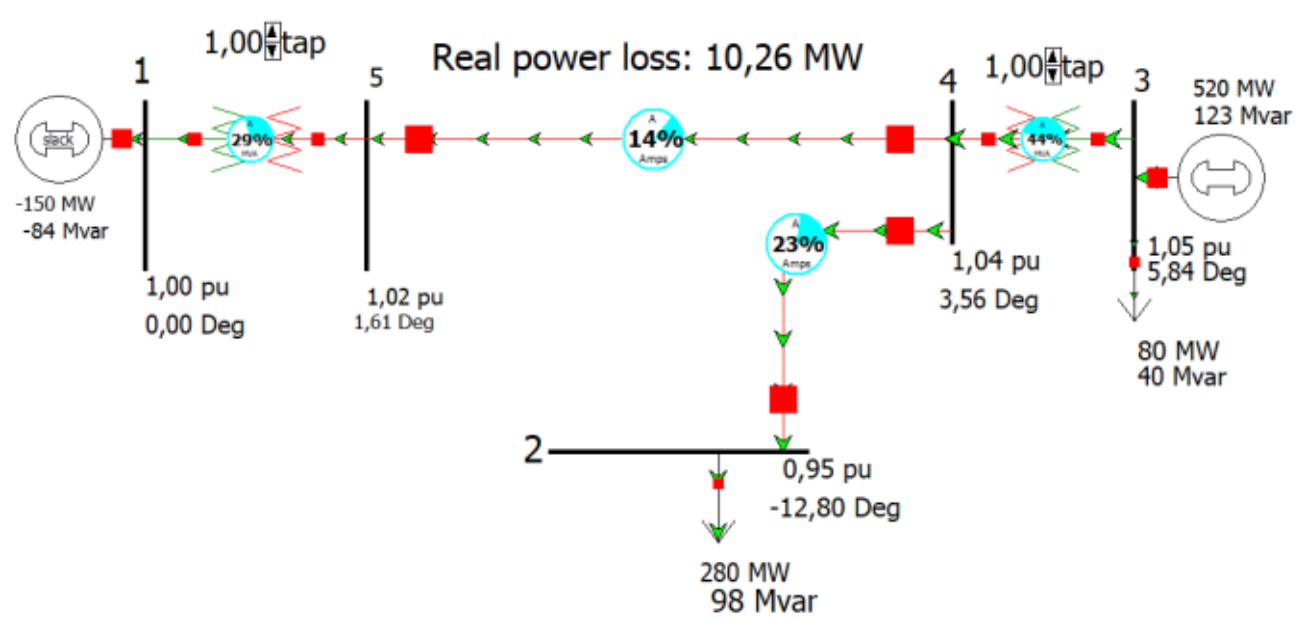

Fig. 23. Simulated 5-bus power system without a transmission line between buses 2 and 5 and shedding loads by 35%.

## *I. Faults Current Calculation on Bus 2*

An essential component of power system analysis is fault analysis, which involves bus voltages and line currents during different kinds of power system failures. Three-phase (3-Ø) balanced faults and unbalanced faults are the two categories of power system faults. Also, there are three different kinds of unbalanced faults, including single line-to-ground (SLG), line-to-line (L-L), and double line-to-ground (DLG) faults.

The definition of the 3-Ø fault is a concurrent short circuit in all three phases. The power systems rarely suffer from this kind of fault. However, its impact on the system is severe when it occurs. The network is solved on a per-phase basis due to its balance. With the exception of the phase shift, the currents in the other phases are equal [2]. Unbalanced faults (L-L, SLG, and DLG) are the most common faults in power systems. In these kinds of faults, the system's operation is only disrupted at the point of the fault. Unbalanced faults have been solved using a variety of techniques. Nonetheless, the approach of symmetrical components, which turns the unbalanced circuit's solution into a solution of many balanced circuits, is employed since the one-line diagram makes the solution of the balanced 3-Ø problems simpler [2]. The fault calculations in this study are performed in PWS based on the following parameters:

- Two generators: $X'' = X^- = 0.12$, $X^o = 0.05$, Y grounded and $Z_n = 0$.
- Two transformers: $Z^o = Z^+ = Z^o$, Y grounded - Y grounded, and $Z_n = 0$.
- Transmission lines: $Z^o = 3Z^+$. The shunt charging (B) is similar to the zero sequence.

The fault calculation results of 3-Ø balanced, SLG, L-L, and DLG faults, using PWS, are presented in Figs. 24 – 27. The 3-Ø balanced fault assumes that phases A, B, and C are short-circuited, SLG assumes a fault between phase A and the ground, L-L faults assume that phases B and C are faulted, and DLG assumes that phases B and C are faulted to the ground. As shown in the fault calculations, the voltages and angles of all the phases impacted by the faults are very low, compared to the phases that are not impacted.

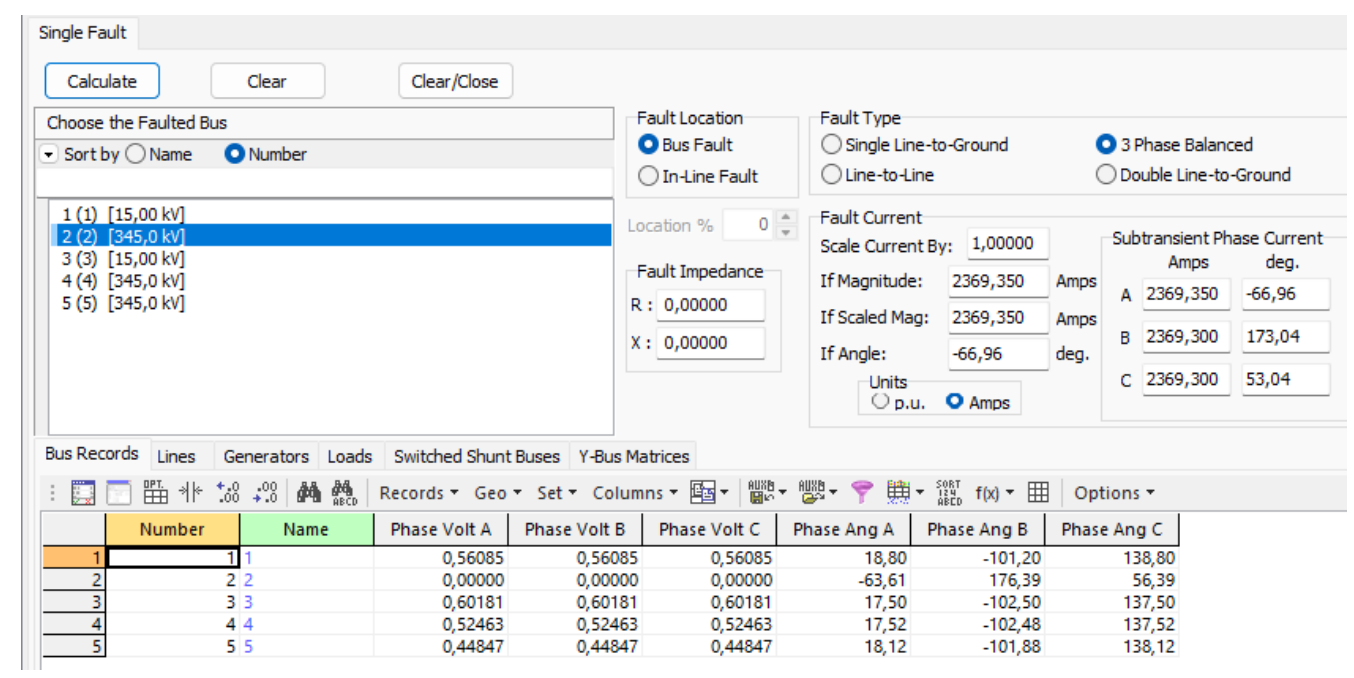

Fig. 24. The calculation results of the 3-Ø balanced faults

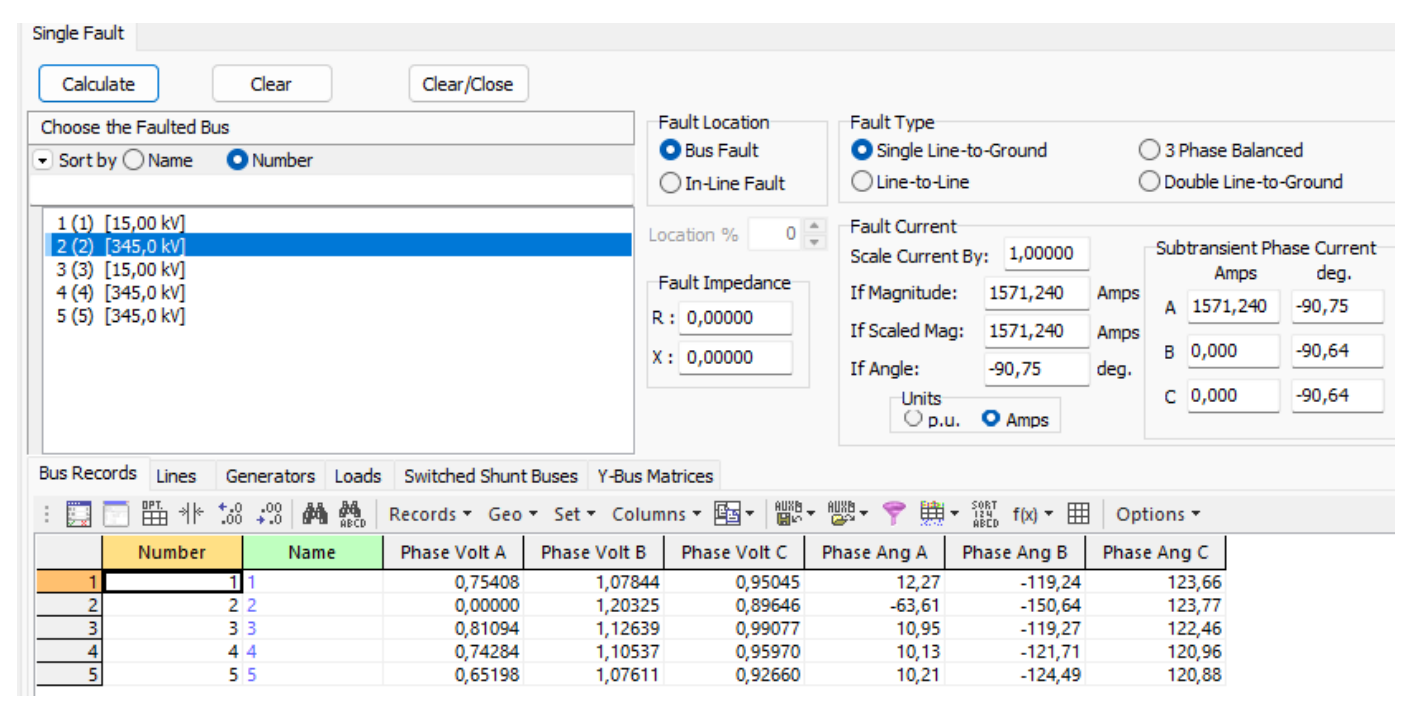

Fig. 25. The calculation results SLG faults

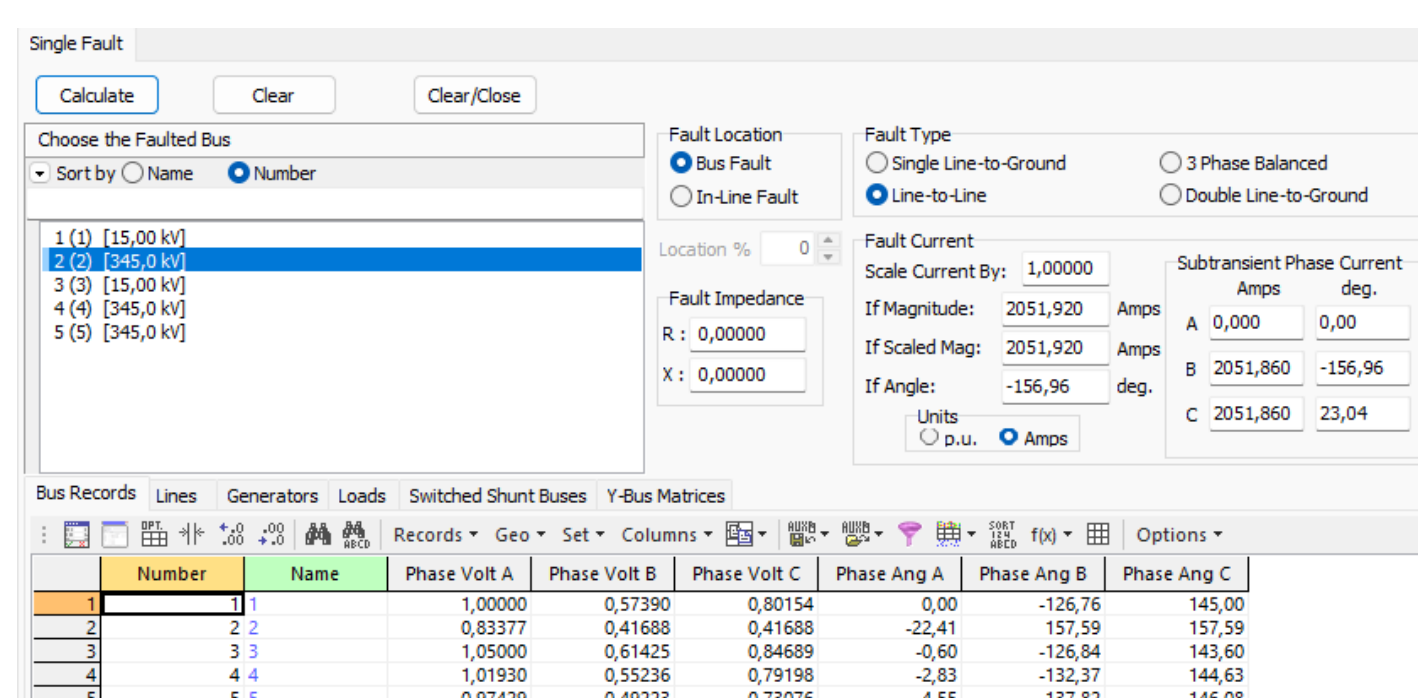

Fig. 26. The calculation results of L-L faults

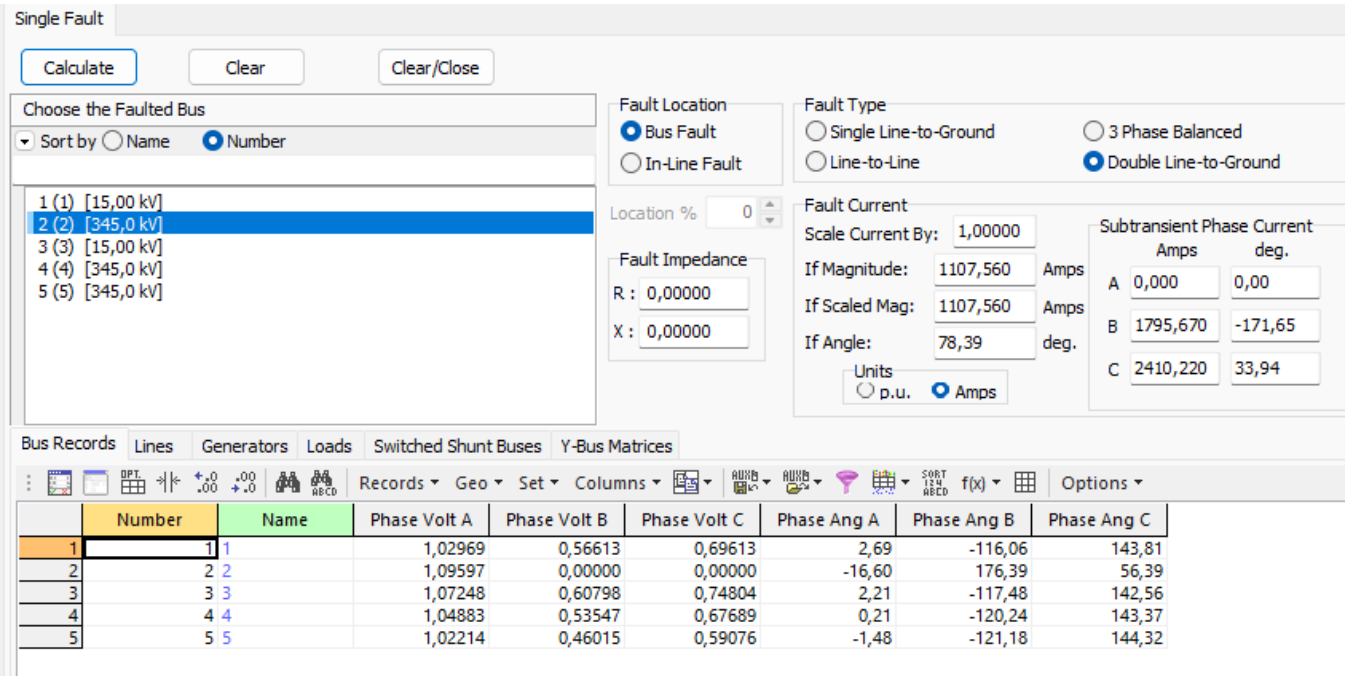

Fig. 27. The calculation results of DLG faults

### J. *Selection of Circuit Breaker*

Proper relay configuration and coordination are made possible by the information gathered from the fault studies. Ground relays employ the L-L fault information, whereas phase relays are chosen and configured using the 3-Ø balanced fault information. The generators' internal impedance and the interruption circuit's impedance determine how much the fault currents will be. The primary emphasis of this study is the rating of the protective switchgear/circuit breakers, which is also determined through fault studies [2].

A mechanical switch that can both interrupt fault currents and close circuits is called a circuit breaker. Certain circuit breakers have the potential to automatically close at a fast speed. The premise behind reclosing is that if a circuit is briefly de-energized, whatever caused the fault has probably broken down, and the ionized arc in the fault has dissipated, as most faults are transient and self-clearing [3].

Based on the fault current calculations, the circuit breakers recommended for the various fault types at Bus 2 are presented in Table 6. These selections are intended to ensure reliable protection against the evaluated fault scenarios.

Table 6. Recommended circuit breakers for different fault currents

| Types of Faults | Fault Currents (A) | Recommended Circuit Breakers |
|---|---|---|
| 3-Ø balanced | 2369.35 | ABB SACE Emax 2 E2.2H/E9 2500 A |
| L-L | 2051.92 | ABB E2.2H/E9 2500 A |
| SLG | 1571.24 | ABB (E2.2N/MS 1600 A |
| DLG | 1107.56 | ABB E1.2N/E9 1250 |

## V. Conclusion

This study presents a power flow analysis of a 5-bus power system using the Newton-Raphson (NR) method. The NR-based power flow analysis offers a robust mathematical framework for estimating active and reactive power flows, phase angles, bus voltages, transformer tap settings, and other steady-state electrical parameters under various operating conditions. In this study, the primary objective was to evaluate system performance under normal operation, transformer tap-changing scenarios, and fault conditions at different buses.

The power system was simulated in PWS for 50 iterations across each operating condition. Simulation results demonstrated that the NR method is effective in capturing the system's steady-state behavior across diverse scenarios. In the base case, the simulation showed that all bus voltages were within acceptable limits except for Bus 2, which violated voltage requirements. This problem was addressed by introducing a shunt capacitor bank at Bus 2 and adjusting the transformer tap ratio, both of which successfully increased the voltage at Bus 2 to acceptable levels. To validate the results, the base case was also simulated using MATLAB. The comparison indicated that PWS provided slightly more accurate and reliable results in terms of power flow calculations.

Further system improvement was explored by introducing a parallel transformer between Buses 1 and 5 and an additional transmission line between Buses 2 and 4. These additions reduced system loading and improved power flow distribution, but at the expense of increased infrastructure and operational costs.

The system was also analyzed under a maintenance scenario, where the transmission line between Buses 2 and 5 was taken out of service. This led to system instability, including a blackout condition, increased real power losses, and a significant voltage drop at Bus 2. An attempt to restore voltage by adding a shunt capacitor at Bus 2 proved insufficient. However, load shedding at Bus 2 by 35% was found to be effective in maintaining voltage levels above 0.95 pu.

Fault analysis, a critical aspect of power system studies, was also conducted. The study evaluated fault currents for various types of faults, including three-phase balanced (3-Ø), single line-to-ground (SLG), line-to-line (L-L), and double line-to-ground (DLG) faults. Based on these calculations, appropriate circuit breaker ratings were recommended for transmission lines between Buses 5–2 and 4–2 to ensure reliable fault protection.